\documentclass[12pt]{article}
\usepackage[left=2.5cm,top=2.50cm,right=2.5cm,bottom=2.50cm]{geometry}
\usepackage{mathrsfs}
\usepackage{amsmath,amssymb,latexsym,color,cancel,graphicx,bbm,colortbl}
\usepackage[english]{babel}
\usepackage[latin1]{inputenc}
\usepackage{ragged2e}
\usepackage{cite}
\begin{document}
\date{}

\title{Algebraic approach for the one-dimensional Dirac-Dunkl oscillator}
\author{D. Ojeda-Guill\'en$^{a}$\footnote{{\it E-mail address:} dojedag@ipn.mx}, R. D. Mota$^{b}$,\\M. Salazar-Ram\'irez$^{a}$ and V. D. Granados$^{c}$} \maketitle

\begin{minipage}{0.9\textwidth}
\small $^{a}$ Escuela Superior de C\'omputo, Instituto Polit\'ecnico Nacional,
Av. Juan de Dios B\'atiz esq. Av. Miguel Oth\'on de Mendiz\'abal, Col. Lindavista,
Delegaci\'on Gustavo A. Madero, C.P. 07738, Ciudad de M\'exico, Mexico.\\

\small $^{b}$ Escuela Superior de Ingenier{\'i}a Mec\'anica y El\'ectrica, Unidad Culhuac\'an,
Instituto Polit\'ecnico Nacional, Av. Santa Ana No. 1000, Col. San Francisco Culhuac\'an, Delegaci\'on Coyoac\'an, C.P. 04430, Ciudad de M\'exico, Mexico.\\

\small $^{c}$ Escuela Superior de F{\'i}sica y Matem\'aticas,
Instituto Polit\'ecnico Nacional, Ed. 9, Unidad Profesional Adolfo L\'opez Mateos, Delegaci\'on Gustavo A. Madero, C.P. 07738, Ciudad de M\'exico, Mexico.\\

\end{minipage}

\begin{abstract}
We extend the $(1+1)$-dimensional Dirac-Moshinsky oscillator by changing the standard derivative by the Dunkl derivative. We demonstrate in a general way that for the Dirac-Dunkl oscillator be parity invariant, one of the spinor component must be even, and the other spinor component must be odd, and vice versa. We decouple the differential equations for each of the spinor component and introduce an appropriate $su(1,1)$ algebraic realization for the cases when one of these functions is even and the other function is odd. The eigenfunctions and the energy spectrum are obtained by using the $su(1,1)$ irreducible representation theory. Finally, by setting the Dunkl parameter to vanish, we show that our results reduce to those of the standard Dirac-Moshinsky oscillator.
\end{abstract}

PACS: 02.20.Sv, 02.30.Jr, 03.65.Fd, 03.65.Pm, \\
Keywords: Dirac equation, Dirac-Moshinsky oscillator, Dunkl derivative, Lie algebras

\section{Introduction}

The Dirac oscillator is a problem that was first studied by Ito \emph{et al.} \cite{Ito} and Cook \cite{Cook}, who added the linear term $-imc\omega\beta{\mathbf{\alpha}}\cdot \mathbf{r}$ to the relativistic momentum $\textbf{p}$ of the free-particle Dirac equation. However, it was until the reintroduction that Moshinsky \cite{Mos,Mos2} made of this problem that it caught the attention it has today. One of the peculiarities that this problem possesses is that in the non-relativistic limit, it reduces to the harmonic oscillator plus a spin-orbit coupling term. The Dirac-Moshinsky oscillator has been extensively studied and applied in many branches of physics, as it can be seen in Refs. \cite{Delgado,Sadurni,Ferkous,Villalba,Benitez,Moreno,deLima}. This problem became relevant in quantum optics when its connection to the Jaynes-Cummings model \cite{Jaynes} was demonstrated \cite{Sadurni,Bermudez,Mandal}.
In particular, in Ref. \cite{NOS0} we have studied algebraically the $(2+1)$-dimensional Dirac-Moshinsky oscillator coupled to an external magnetic field and computed its relativistic coherent states. Schr\"odinger introduced coherent states in the late 1920s as the most classical ones of the harmonic oscillator \cite{Sch} and Glauber reintroduced these states into the field of quantum optics in 1963 \cite{Glau}. Despite this fact, coherent states are still widely studied today in many branches of physics, as can be seen from the references \cite{Israel,Bozzio,Marwah,Chung2,Chung3,Chung4}.

On the other hand, the Dunkl operators $D_{x_i}$ were introduced in the context of the deformed oscillator by Yang \cite{YANG}. These operators are combinations of differential and difference operators, associated to a finite reflection group $\mathcal{G}$. As in the case of the Dirac oscillator problem, Dunkl reintroduced these operators to study polynomials in several variables with reflection symmetries related to finite reflection groups \cite{DUNKL}. In this context, the Dunkl Laplacian is a combination of the classical Laplacian in $\mathbb{R}^n$ with some difference terms, such that the resulting operator is only invariant under $\mathcal{G}$ and it is not under the whole orthogonal group \cite{DUNKL}. Moreover, the Dunkl operators are central to the theory of multivariate orthogonal polynomials \cite{XuY} and have been applied to harmonic analysis \cite{Rosler}.

The Dunkl operators have been used to study Hamiltonians with reflection operators, which most notably appear in the context of the Calogero-Sutherland integrable models and their generalizations \cite{Brink,Lapointe}. Later, these operators were used to study some problems of quantum mechanics such as isotropic oscillator and the Coulomb problem \cite{GEN1,GEN2,GEN3,GEN4}. In these works, the authors obtained the symmetry of each problem (called Schwinger-Dunkl algebra) and the exact solutions of the Schr\"odinger equation in terms of Jacobi, Laguerre and Hermite polynomials. In fact, in Refs. \cite{NOS1,NOS2} we studied the two-dimensional Dunkl-oscillator and the Dunkl-Coulomb problems in terms of the $su(1,1)$ Lie algebra and the theory of unitary irreducible representations.

Recently, the Dunkl derivatives were also used to study some non-relativistic and relativistic problems, such as the one-dimensional box problem \cite{Chung5}, the Dirac equation and the Dirac harmonic oscillator in one and two dimensions \cite{Chung,NOS3}. The aim of the present work is to introduce an $SU(1,1)$ algebraic approach to obtain the eigenfunctions and energy spectrum of the $(1+1)$-dimensional Dirac-Dunkl oscillator.

This work is organized as follows. In Section $2$, the main properties of the Dirac-Moshinsky oscillator in $(1+1)$ dimensions are revisited. Then, we introduce the Dunkl derivative to obtain what we called the Dirac-Dunkl oscillator and the differential equations of the spinor components. As an application of the one-dimensional Dirac-Dunkl-Moshinsky oscillator, we are able to define the Anti-Jaynes-Cummings Hamiltonian model of quantum optics. In Section $3$ we study the standard Dirac-Moshinsky oscillator and the Dirac-Dunkl oscillator under space reflections. We show that the Dirac equation for both problems is invariant under parity if and only if one of the spinor components is an even function and the other is an odd function. In Section $4$, we give two different realizations and the eigenfunctions basis (Sturmian basis) for an irreducible unitary representations of the $su(1,1)$ Lie algebra. With these $su(1,1)$ realizations, we study the cases in which one the spinor components is an even function and the other spinor component is an odd function to obtain the energy spectrum and eigenfunctions for each case. The Section $5$ is dedicated to construct the $SU(1,1)$ Perelomov coherent states of the Dirac-Dunkl oscillator and to recovering the standard Dirac-Moshinsky oscillator by setting the Dunkl parameter to vanish ($\mu=0$). Finally, we give some concluding remarks.

\section{The Dirac oscillator and the Dirac-Dunkl oscillator in one dimension}

In $(1+1)$ dimensions, the time-independent Dirac equation for the Dirac-Moshinsky oscillator is given by the Hamiltonian \cite{Mos}
\begin{equation}
H_D\Psi=\left[c \mathbf{\alpha} \cdot\left(\mathbf{p}-im\omega \mathbf{r}\beta\right)+mc^2\beta\right]\Psi=E\Psi.
\end{equation}
We shall consider the most convenient representations of the Dirac matrices $\alpha$ and $\beta$
\begin{equation}
\alpha=\begin{pmatrix}
0 & -i \\
i & 0 \end{pmatrix},\quad\quad \beta=\begin{pmatrix}
1 & 0 \\
0 & -1 \end{pmatrix},
\end{equation}
which satisfy the Clifford algebra
\begin{align}\nonumber
\alpha_a\alpha_b+\alpha_b\alpha_a=2\delta_{ab}\textbf{1},\\
\alpha_a\beta+\beta\alpha_a=0,\\\nonumber
\alpha_a^2=\beta^2=\texttt{1}.
\end{align}

With this realization of the Dirac matrices we obtain the following coupled equations
\begin{equation}
\left(E-mc^2\right)|\Psi_1\rangle=c(-ip_x+m\omega x)|\Psi_2\rangle,\label{c1}
\end{equation}
\begin{equation}
\left(E+mc^2\right)|\Psi_2\rangle=c(ip_x+m\omega x)|\Psi_1\rangle.\label{c2}
\end{equation}
If we introduce the usual creation and annihilation operators
\begin{equation}
a=\sqrt{\frac{m\omega}{2\hbar}}x+\frac{i}{\sqrt{2m\omega\hbar}}p_x,\quad\quad
a^{\dag}=\sqrt{\frac{m\omega}{2\hbar}}x-\frac{i}{\sqrt{2m\omega\hbar}}p_x,
\end{equation}
these coupled equations can be written in the simplified form \cite{Toyama}
\begin{equation}
\left(E-mc^2\right)|\Psi_1\rangle=c\sqrt{2m\omega\hbar}a^{\dag}|\Psi_2\rangle,\label{c3}
\end{equation}
\begin{equation}
\left(E+mc^2\right)|\Psi_2\rangle=c\sqrt{2m\omega\hbar}a|\Psi_1\rangle.\label{c4}
\end{equation}
The uncoupled equations for the spinor components $\Psi_1$ and $\Psi_2$ are given by
\begin{equation}
2c^2m\omega\hbar aa^{\dag}\Psi_2=\left(E^2-m^2c^4\right)\Psi_2,\label{ep2}
\end{equation}
\begin{equation}
2c^2m\omega\hbar a^{\dag}a\Psi_1=\left(E^2-m^2c^4\right)\Psi_1.\label{ep1}
\end{equation}

These equations can be solved by using the non-relativistic theory of the quantum harmonic oscillator. Hence, the eigenvalues of the one-dimensional Dirac-Moshinsky oscillator are \cite {Nogami,Rados}
\begin{equation}
E_n=\pm mc^2\sqrt{1+\frac{2|n|\hbar\omega}{mc^2}}, \quad\quad n=0,\pm1,\pm2,...\label{spectrum1}
\end{equation}
where the upper sign should be chosen for $n\geq0$ and the lower one for $n<0$. The normalized
eigenfunction $|\Psi\rangle$ is
\begin{equation}
|\Psi\rangle=\begin{pmatrix}
\sqrt{\frac{\lambda(E+mc^2)}{2^{|n|+1}|n|!\sqrt{\pi}E}}H_{|n|}(\lambda x)e^{-\lambda^2x^2/2}\\
\sqrt{\frac{\lambda(E-mc^2)}{2^{|n|}(|n|-1)!\sqrt{\pi}E}}H_{|n|-1}(\lambda x)e^{-\lambda^2x^2/2}
\end{pmatrix},\label{eigen1}
\end{equation}
where $H_n(x)$ is the Hermite polynomial and $\lambda=\sqrt{\frac{m\omega}{\hbar}}$.

Now, we introduce the Dunkl derivative $D_x^{\mu}$ defined by
\begin{equation}
D_x^{\mu}=\frac{d}{d x}+\frac{\mu}{x}\left(\mathbb{I}-R\right),
\end{equation}
where $R$ is the reflection operator with respect to the plane $x=0$ and produces the following action $Rf(x)=f(-x)$. With the Dunkl derivative we can define the Dunkl momentum operator $p_x^{\mu}$ as $p_x^{\mu}=-i\hbar D_x^{\mu}$ and the generalized-Dunkl creation $a^{\dag}_D$ and annihilation $a_D$ operators as
\begin{equation}
a^{\dag}_D=\frac{1}{\sqrt{2m\omega\hbar}}\left[m\omega x-\hbar\left(\frac{d}{d x}+\frac{\mu}{x}\left(\mathbb{I}-R\right)\right)\right],\label{a+D}
\end{equation}
\begin{equation}
a_D=\frac{1}{\sqrt{2m\omega\hbar}}\left[m\omega x+\hbar\left(\frac{d}{d x}+\frac{\mu}{x}\left(\mathbb{I}-R\right)\right)\right].\label{aD}
\end{equation}
If we substitute these definitions into equations (\ref{c3}) and (\ref{c4}), it can be shown that the Hamiltonian of the one-dimensional Dirac-Dunkl oscillator can be mapped onto the Anti-Jaynes-Cummings model of quantum optics as
\begin{equation}
H=\delta\left(\sigma_-a_D+\sigma_+a^{\dag}_D\right)+mc^2\sigma_z,
\end{equation}
where $\delta=c\sqrt{2m\omega\hbar}$, $\sigma_+$ and $\sigma_-$ are the spin raising and lowering operators, and $\sigma_z$ is the Pauli matrix. Also, from expressions (\ref{a+D}) and (\ref{aD}) it follows that
\begin{equation}
a_Da_D^{\dag}=\frac{1}{2m\hbar\omega}\left[m^2\omega^2x^2+2m\omega\hbar\mu R+m\omega\hbar-\hbar^2\left(\frac{d^2}{d x^2}+2\frac{\mu}{x}\frac{d}{d x}-\frac{\mu}{x^2}(1-R)\right)\right],
\end{equation}
\begin{equation}
a_D^{\dag}a_D=\frac{1}{2m\hbar\omega}\left[m^2\omega^2x^2-2m\omega\hbar\mu R-m\omega\hbar-\hbar^2\left(\frac{d^2}{d x^2}+2\frac{\mu}{x}\frac{d}{d x}-\frac{\mu}{x^2}(1-R)\right)\right].
\end{equation}
Therefore, these new creation $a^{\dag}_D$ and annihilation $a_D$ operators satisfy
\begin{equation}
[a_D,a^{\dag}_D]=1+2\mu R.
\end{equation}

From equations (\ref{ep2}) and (\ref{ep1}) we obtain that the uncoupled equations of the spinor components $\Psi_1$ and $\Psi_2$ for the $(1+1)$ Dirac-Dunkl oscillator are given by
\begin{equation}
\left[m^2\omega^2x^2-2m\omega\hbar\mu R-m\omega\hbar-\hbar^2\left(\frac{d^2}{d x^2}+2\frac{\mu}{x}\frac{d}{d x}-\frac{\mu}{x^2}(1-R)\right)\right]\Psi_1=\left[\left(\frac{E}{c}\right)^2-m^2c^2\right]\Psi_1,\label{e1}
\end{equation}
\begin{equation}
\left[m^2\omega^2x^2+2m\omega\hbar\mu R+m\omega\hbar-\hbar^2\left(\frac{d^2}{d x^2}+2\frac{\mu}{x}\frac{d}{d x}-\frac{\mu}{x^2}(1-R)\right)\right]\Psi_2=\left[\left(\frac{E}{c}\right)^2-m^2c^2\right]\Psi_2.\label{e2}
\end{equation}
However, at this stage it is not clear the symmetry under parity that we must impose on the eigenfunctions $\Psi_1$ and $\Psi_2$ to solve these equations. This issue will be clarified in the next Section.

\section{Space reflections in the $(1+1)$-Dirac-Dunkl oscillator}

In this Section we shall study the symmetry conditions under space reflections that the spinor components must satisfy so that the complete Dirac-Dunkl equation is invariant under the parity operator. To do this, we fist apply the parity operator $R$ to the standard Dirac-Moshinsky oscillator in $(1+1)$ dimensions
\begin{equation}
\left[c \mathbf{\alpha} \left(Rp_x-im\omega Rx\beta\right)+mc^2\beta R\right]\Psi(x,t)=ER\Psi(x,t).
\end{equation}
Using that $Rp_x=-p_xR$, $Rx=-xR$, and $R\Psi(x,t)=\Psi(-x,t)$ we obtain the following expression
\begin{equation}
\left[-c \mathbf{\alpha} \left(p_x-im\omega x\beta\right)+mc^2\beta \right]\Psi(-x,t)=E\Psi(-x,t).\label{pdm}
\end{equation}
Now, if we define the operator $\mathcal{P}=e^{i\varphi}\beta$, then $\mathcal{P}^{-1}=e^{-i\varphi}\beta$. Thus, we obtain that $\mathcal{P}^{-1}\alpha\mathcal{P}=-\alpha$. Moreover, since the operator $\mathcal{P}$ satisfies $[\mathcal{P},\beta]=0$, $[\mathcal{P},p_x]=0$, and $[\mathcal{P},x]=0$, we can write the equation (\ref{pdm}) as
\begin{equation}
\left[c \mathcal{P}^{-1}\mathbf{\alpha} \left(p_x-im\omega x\beta\right)+mc^2e^{-i\varphi}\right]\mathcal{P}\Psi(-x,t)=E\Psi(-x,t).
\end{equation}
After multiplying this equation on the left by the operator $\mathcal{P}$ we obtain that
\begin{equation}
\left[c \mathbf{\alpha} \left(p_x-im\omega x\beta\right)+mc^2\beta\right]\mathcal{P}\Psi(-x,t)=E\mathcal{P}\Psi(-x,t).
\end{equation}
Therefore, the Dirac equation of the $(1+1)$ Dirac-Moshinsky oscillator in invariant under parity if and only if $\Psi(x,t)=\mathcal{P}\Psi(-x,t)$. Explicitly we have
\begin{equation}
\begin{pmatrix}
\Psi_1(x,t)\\
\Psi_2(x,t)
\end{pmatrix}=
\begin{pmatrix}
e^{i\varphi}\Psi_1(-x,t)\\
-e^{-i\varphi}\Psi_2(-x,t)
\end{pmatrix}.
\end{equation}
In particular, if we set the phase $\varphi=0$ we conclude that $\Psi_1$ must be an even function and $\Psi_2$ an odd function. Similarly, if $\varphi=\pi$ we have that $\Psi_1$ must be an odd function and $\Psi_2$ an even function.

Now, we apply this procedure to the $(1+1)$ Dirac-Dunkl oscillator. From the previous definition of the Dunkl momentum operator $p_x^{\mu}$, we obtain that $Rp_x^{\mu}=-p_x^{\mu}R$. Thus, from this result and by applying the parity operator $R$ to the $(1+1)$ Dirac-Dunkl oscillator, we obtain
\begin{equation}
\left[-c \mathbf{\alpha} \left(p_x^{\mu}-im\omega x\beta\right)+mc^2\beta \right]\Psi(-x,t)=E\Psi(-x,t).\label{pddm}
\end{equation}
Since $\mathcal{P}$ commutes with $\beta$, $p_x^{\mu}$, and $x$, and $\mathcal{P}^{-1}\alpha\mathcal{P}=-\alpha$, we can write this equation as
\begin{equation}
\left[c \mathcal{P}^{-1}\mathbf{\alpha} \left(p_x^{\mu}-im\omega x\beta\right)+mc^2e^{-i\varphi}\right]\mathcal{P}\Psi(-x,t)=E\Psi(-x,t).
\end{equation}
After multiplying this equation on the left by $\mathcal{P}$ we obtain the following result
\begin{equation}
\left[c \mathbf{\alpha} \left(p_x^{\mu}-im\omega x\beta\right)+mc^2\beta\right]\mathcal{P}\Psi(-x,t)=E\mathcal{P}\Psi(-x,t).
\end{equation}
Thus, the $(1+1)$ Dirac-Dunkl oscillator in invariant under parity if and only if $\Psi(x,t)=\mathcal{P}\Psi(-x,t)$ and the parity of the spinor components are well defined. Similarly to the case of the Dirac-Moshinsky oscillator, if we set $\Psi_1$ as an even function, this implies that $\Psi_2$ must be set as an odd function and vice versa.

\section{$SU(1,1)$ approach for the $(1+1)$ Dirac-Dunkl oscillator}

From the results obtained in the previous Sections, we shall study the problem of the $(1+1)$ Dirac-Dunkl oscillator using the $SU(1,1)$ group theory
for two cases:  i) when the spinor component $\Psi_1$ is an even function and $\Psi_2$ is an odd function, and ii) when the spinor component $\Psi_1$ is an odd function and $\Psi_2$ is an even function.

The $su(1,1)$ Lie algebra is spanned by the generators $K_{+}$, $K_{-}$ and $K_{0}$, which satisfy the commutation relations \cite{Vourdas}
\begin{eqnarray}
[K_{0},K_{\pm}]=\pm K_{\pm},\quad\quad [K_{-},K_{+}]=2K_{0}.\label{com}
\end{eqnarray}
The action of these operators on the Sturmian basis $\{|k,n\rangle, n=0,1,2,...\}$ is
\begin{equation}
K_{+}|k,n\rangle=\sqrt{(n+1)(2k+n)}|k,n+1\rangle,\label{k+n}
\end{equation}
\begin{equation}
K_{-}|k,n\rangle=\sqrt{n(2k+n-1)}|k,n-1\rangle,\label{k-n}
\end{equation}
\begin{equation}
K_{0}|k,n\rangle=(k+n)|k,n\rangle,\label{k0n}
\end{equation}
where $|k,0\rangle$ is the lowest normalized state. The Casimir
operator for any irreducible representation satisfies
\begin{equation}
K^{2}=-K_{+}K_{-}+K_{0}(K_{0}-1)=k(k-1).\label{cas}
\end{equation}

In this sense, we can introduce the following realization of the $su(1,1)$ Lie algebra
\begin{equation}
K_{0}^{+}=\frac{1}{4}\left[-\frac{d^2}{dr^2}-\frac{2\mu}{r}\frac{d}{dr}+r^2\right],\label{rk01}
\end{equation}
\begin{equation}
K_{+}^{+}=\frac{1}{2}\left[r\frac{d}{dr}-r^2+2K_{0}^{+}+\left(\frac{1}{2}+\mu\right)\right],\label{rk+1}
\end{equation}
\begin{equation}
K_{-}^{+}=\frac{1}{2}\left[-r\frac{d}{dr}-r^2+2K_{0}^{+}-\left(\frac{1}{2}+\mu\right)\right].\label{rk-1}
\end{equation}
A direct computation shows that the Casimir operator for these operators is given by
\begin{equation}
K^{2,+}=\frac{1}{16}\left(4\mu^2-4\mu-3\right)=k(k-1),
\end{equation}
and the Bargmann index $k$ for this realization takes the following values
\begin{equation}
k_1^+=\frac{1}{4}+\frac{1}{2}\mu, \quad\quad\quad\quad k_2^+=\frac{3}{4}-\frac{1}{2}\mu.
\end{equation}
This realization will be useful in the case where the spinor component is an even function. For the case when the spinor component is an odd function, we can use the following operators
\begin{equation}
K_{0}^{-}=\frac{1}{4}\left[-\frac{d^2}{dr^2}-\frac{2\mu}{r}\frac{d}{dr}+\frac{2\mu}{r^2}+r^2\right],\label{rk02}
\end{equation}
\begin{equation}
K_{+}^{-}=\frac{1}{2}\left[r\frac{d}{dr}-r^2+2K_{0}^{-}+\left(\frac{1}{2}+\mu\right)\right],\label{rk+2}
\end{equation}
\begin{equation}
K_{-}^{-}=\frac{1}{2}\left[-r\frac{d}{dr}-r^2+2K_{0}^{-}-\left(\frac{1}{2}+\mu\right)\right].\label{rk-2}
\end{equation}
From this realization we obtain that the Casimir operator is
\begin{equation}
K^{2,-}=\frac{1}{16}\left(4\mu^2+4\mu-3\right)=k(k-1),
\end{equation}
and the Bargmann index is given by
\begin{equation}
k_1^{-}=\frac{1}{4}-\frac{1}{2}\mu, \quad\quad\quad\quad k_2^{-}=\frac{3}{4}+\frac{1}{2}\mu.
\end{equation}

Therefore, since we are interested only in the positive discrete representations of the $SU(1,1)$ group, for which
$k>0$, we will only consider the values
\begin{equation}
k_1^{+}=\frac{1}{4}+\frac{1}{2}\mu, \quad\quad\quad\quad k_2^{-}=\frac{3}{4}+\frac{1}{2}\mu.
\end{equation}

The eigenfunctions basis for an irreducible unitary representations of the $su(1,1)$ Lie algebra (Sturmian basis) in terms of the quantum group numbers $n$, $k$ are
\begin{equation}
R_n^k(r)=\left[\frac{2\Gamma(n+1)}{\Gamma(n+2k)}\right]^{1/2}r^{2k-\frac{1}{2}}e^{-r^2/2}L_n^{2k-1}(r^2),\label{gen}
\end{equation}
where $L_n^{2k-1}(r^2)$ are the associated Laguerre polynomials. It is important to note that this Sturmian basis is used for harmonic oscillator problems \cite{Gerry,Gur}.

\subsection{$\Psi_1$ even function and $\Psi_2$ odd function}

Now, we shall consider that the spinor component $\Psi_1$ is and even function $(R=1)$ and $\Psi_2$ is an odd function $(R=-1)$. Then, from equations
(\ref{e1}) and (\ref{e2}) we obtain the following uncoupled equations
\begin{equation}
\left[m^2\omega^2x^2-2m\omega\hbar\mu -m\omega\hbar-\hbar^2\frac{d^2}{d x^2}-\frac{2\mu\hbar^2}{x}\frac{d}{d x}\right]\Psi_1(r)=\left[\left(\frac{E}{c}\right)^2-m^2c^2\right]\Psi_1(r),
\end{equation}
\begin{equation}
\left[m^2\omega^2x^2-2m\omega\hbar\mu +m\omega\hbar-\hbar^2\frac{d^2}{d x^2}-\frac{2\mu\hbar^2}{x}\frac{d}{d x}+\frac{2\mu\hbar^2}{x^2}\right]\Psi_2(r)=\left[\left(\frac{E}{c}\right)^2-m^2c^2\right]\Psi_2(r).
\end{equation}
If we introduce the variable $x=br$, with $b=\sqrt{\frac{\hbar}{m\omega}}$ we can write these differential equations as
\begin{equation}
\left[-\frac{d^2}{dr^2}-\frac{2\mu}{r}\frac{d}{dr}+r^2-(1+2\mu)-\frac{E^2}{\hbar\omega mc^2}+\frac{mc^2}{\hbar\omega}\right]\Psi_1(r)=0,\label{epr1}
\end{equation}
\begin{equation}
\left[-\frac{d^2}{dr^2}-\frac{2\mu}{r}\frac{d}{dr}+\frac{2\mu}{r^2}+r^2+(1-2\mu)-\frac{E^2}{\hbar\omega mc^2}+\frac{mc^2}{\hbar\omega}\right]\Psi_2(r)=0.\label{eir2}
\end{equation}
Since the spinor component $\Psi_1$ is even we will consider the value of $k_1^+$. Hence, from the operator (\ref{rk01}), we can write the equation (\ref{epr1}) for $\Psi_1$ as follows
\begin{equation}
\left[K_{0}^{+}-\left(\frac{1}{4}+\frac{1}{2}\mu\right)-\frac{E^2}{4\hbar\omega mc^2}+\frac{mc^2}{4\hbar\omega}\right]\Psi_1(r)=0.
\end{equation}
Thus, from equation (\ref{k0n}) we obtain for the spinor component $\Psi_1$ that
\begin{equation}
K_{0}^{+}\Psi_1(r)=\left[\left(\frac{1}{4}+\frac{1}{2}\mu\right)+\frac{E^2}{4\hbar\omega mc^2}-\frac{mc^2}{4\hbar\omega}\right]\Psi_1(r)=(n+k_1^+)\Psi_1(r),
\end{equation}
Therefore, from the fact that $k_1^+=\frac{1}{4}+\frac{1}{2}\mu$, we can obtain the energy spectrum $E$ in terms of the quantum number $n$
\begin{equation}
E=\pm mc^2\sqrt{1+\frac{4\hbar\omega}{mc^2}n}, \quad\quad n=0,1,2,...
\end{equation}
Now, the explicit form of $\Psi_1$ can be obtained from equation (\ref{gen})
\begin{equation}
\Psi_1(r)=\left[\frac{2\Gamma(n+1)}{\Gamma(n+\frac{1}{2}+\mu)}\right]^{1/2}r^{\mu}e^{-r^2/2}L_n^{\mu-\frac{1}{2}}(r^2).
\end{equation}

Similarly, since $\Psi_2$ is an odd function, we can use the realization given in equations (\ref{rk02})-(\ref{rk-2}). Thus, the differential equation (\ref{eir2}) for $\Psi_2$ can be written as
\begin{equation}
\left[K_{0}+\left(\frac{1}{4}-\frac{1}{2}\mu\right)-\frac{E^2}{4\hbar\omega mc^2}+\frac{mc^2}{4\hbar\omega}\right]\Psi_2(r)=0,
\end{equation}
and
\begin{equation}
K_{0}\Psi_2(r)=\left[-\left(\frac{1}{4}-\frac{1}{2}\mu\right)+\frac{E^2}{4\hbar\omega mc^2}-\frac{mc^2}{4\hbar\omega}\right]\Psi_1(r)=(n'+k_2^-)\Psi_2(r),
\end{equation}
Therefore, by using that $k_2^{-}=\frac{3}{4}+\frac{1}{2}\mu$, the energy spectrum $E$ is given by
\begin{equation}
E=\pm mc^2\sqrt{1+\frac{4\hbar\omega}{mc^2}(n'+1)}, \quad\quad n=0,1,2,...
\end{equation}
Now, since both eigenfunctions belong to the same energy level, we must change $n'\rightarrow n-1$ and the spinor component $\Psi_2$ is explicitly given by
\begin{equation}
\Psi_2(r)=\left[\frac{2\Gamma(n)}{\Gamma(n+\frac{1}{2}+\mu)}\right]^{1/2}r^{1+\mu}e^{-r^2/2}L_{n-1}^{\mu+\frac{1}{2}}(r^2).
\end{equation}
Therefore, after a general normalization we obtain that the eigenfunction $\Psi(r)$ when $\Psi_1$ is an even function and $\Psi_2$ is an odd function is given by
\begin{equation}
|\Psi\rangle=\left[\frac{2\Gamma(n+1)}{\Gamma(n+\frac{1}{2}+\mu)}\right]^{1/2}r^{\mu}e^{-r^2/2}\begin{pmatrix}
\sqrt{\frac{E\pm mc^2}{2E}}L_n^{\mu-\frac{1}{2}}(r^2)\\
\mp i\sqrt{\frac{E\mp mc^2}{2E}}\frac{r}{\sqrt{n}}L_{n-1}^{\mu+\frac{1}{2}}(r^2)
\end{pmatrix},\label{fin1}
\end{equation}
and its energy spectrum if given by
\begin{equation}
E=\pm mc^2\sqrt{1+\frac{4\hbar\omega}{mc^2}n}, \quad\quad n=0,1,2,...
\end{equation}

Here it is important to point out that, as can be seen in equation (\ref{eir2}), the Dunkl derivative introduces a fictitious angular momentum to the standard Dirac harmonic oscillator and the Dunkl derivative parameter $\mu$ takes the role of the angular momentum eigenvalue (see Refs. \cite{Gerry,Gur}). In addition, the dependence of the solution of the problem on the Dunkl derivative is reflected only in the eigenfunction and not in the energy spectrum.

\subsection{$\Psi_1$ odd function and $\Psi_2$ even function}

Now, we shall consider that $\Psi_1$ is an odd function and $\Psi_2$ is an even function. For this case we obtain from equations (\ref{e1}) and (\ref{e2}) the following differential equations
\begin{equation}
\left[m^2\omega^2x^2+2m\omega\hbar\mu-m\omega\hbar-\hbar^2\frac{d^2}{d x^2}-\frac{2\mu\hbar^2}{x}\frac{d}{d x}+\frac{2\mu\hbar^2}{x^2}\right]\Psi_1(r)=\left[\left(\frac{E}{c}\right)^2-m^2c^2\right]\Psi_1(r),\label{ep1o}
\end{equation}
\begin{equation}
\left[m^2\omega^2x^2+2m\omega\hbar\mu+m\omega\hbar-\hbar^2\frac{d^2}{d x^2}-\frac{2\mu\hbar^2}{x}\frac{d}{d x}\right]\Psi_2(r)=\left[\left(\frac{E}{c}\right)^2-m^2c^2\right]\Psi_2(r).\label{ep2e}
\end{equation}
With the definitions $x=br$, $b=\sqrt{\frac{\hbar}{m\omega}}$ we obtain that
\begin{equation}
\left[-\frac{d^2}{dr^2}-\frac{2\mu}{r}\frac{d}{dr}+\frac{2\mu}{r^2}+r^2-(1-2\mu)-\frac{E^2}{\hbar\omega mc^2}+\frac{mc^2}{\hbar\omega}\right]\Psi_1(r)=0,\label{eir1}
\end{equation}
\begin{equation}
\left[-\frac{d^2}{dr^2}-\frac{2\mu}{r}\frac{d}{dr}+r^2+(1+2\mu)-\frac{E^2}{\hbar\omega mc^2}+\frac{mc^2}{\hbar\omega}\right]\Psi_2(r)=0.\label{epr2}
\end{equation}
In order to obtain the explicit form of $\Psi_1$ and its energy spectrum, we shall use the $su(1,1)$ realization of equations (\ref{rk01})-(\ref{rk-1}). However, since the spinor component $\Psi_1$ is an odd function, the Bargmann index for this case is $k_2^-=\frac{3}{4}+\frac{1}{2}\mu$. Thus, we obtain from equations (\ref{eir1}) and (\ref{rk01}) that
\begin{equation}
K_{0}\Psi_1(r)=\left[\left(\frac{1}{4}-\frac{1}{2}\mu\right)+\frac{E^2}{4\hbar\omega mc^2}-\frac{mc^2}{4\hbar\omega}\right]\Psi_1(r)=(n'+k_2^-)\Psi_1(r),
\end{equation}
Therefore, in this case where $\Psi_1$ is an odd function the energy spectrum $E$ is
\begin{equation}
E=\pm mc^2\sqrt{1+\frac{4\hbar\omega}{mc^2}\left(n'+\frac{1}{2}+\mu\right)}, \quad\quad n=0,1,2,...
\end{equation}
and its Sturmian basis is now explicitly given by
\begin{equation}
\Psi_1(r)=\left[\frac{2\Gamma(n'+1)}{\Gamma(n'+\frac{3}{2}+\mu)}\right]^{1/2}r^{1+\mu}e^{-r^2/2}L_n'^{\mu+\frac{1}{2}}(r^2).
\end{equation}

The differential equation for the spinor component $\Psi_2$ in terms of $r$ is given by
\begin{equation}
\left[-\frac{d^2}{dr^2}-\frac{2\mu}{r}\frac{d}{dr}+r^2+(1+2\mu)-\frac{E^2}{\hbar\omega mc^2}+\frac{mc^2}{\hbar\omega}\right]\Psi_2(r)=0.\label{epr22}
\end{equation}
Since $\Psi_2$ is an even function, we can use the $su(1,1)$ realization of equations (\ref{rk01})-(\ref{rk-1}) with $k_1^+=\frac{1}{4}+\frac{1}{2}\mu$. Thus, from equations (\ref{rk01}) and (\ref{epr22}) we obtain the following result
\begin{equation}
K_{0}\Psi_2(r)=\left[-\left(\frac{1}{4}+\frac{1}{2}\mu\right)+\frac{E^2}{4\hbar\omega mc^2}-\frac{mc^2}{4\hbar\omega}\right]\Psi_2(r)=(n+k_1^+)\Psi_2(r).
\end{equation}
Therefore, the energy spectrum $E$ and its eigenfunction are
\begin{equation}
E=\pm mc^2\sqrt{1+\frac{4\hbar\omega}{mc^2}\left(n+\frac{1}{2}+\mu\right)}, \quad\quad n=0,1,2,...
\end{equation}
\begin{equation}
\Psi_2(r)=\left[\frac{2\Gamma(n+1)}{\Gamma(n+\frac{1}{2}+\mu)}\right]^{1/2}r^{\mu}e^{-r^2/2}L_n^{\mu-\frac{1}{2}}(r^2).
\end{equation}
If we impose that both eigenfunctions belong to the same energy level, then $n'\rightarrow n-1$. After a general normalization we obtain that the eigenfunction $\Psi(r)$ for which $\Psi_1$ is an odd function, and $\Psi_2$ is an even function is given by
\begin{equation}
|\Psi\rangle=\left[\frac{2\Gamma(n+1)}{\Gamma(n+\frac{1}{2}+\mu)}\right]^{1/2}r^{\mu}e^{-r^2/2}\begin{pmatrix}
\sqrt{\frac{E\pm mc^2}{2E}}\frac{r}{\sqrt{n}}L_{n-1}^{\mu+\frac{1}{2}}(r^2)\\
\mp i\sqrt{\frac{E\mp mc^2}{2E}}L_n^{\mu-\frac{1}{2}}(r^2)
\end{pmatrix},\label{fin2}
\end{equation}
and its energy spectrum is given by
\begin{equation}
E=\pm mc^2\sqrt{1+\frac{4\hbar\omega}{mc^2}\left(n+\frac{1}{2}+\mu\right)}, \quad\quad n=0,1,2,...
\end{equation}

Also, for this case, we observe from equation (\ref{eir1}) that the Dunkl derivative again introduces a fictitious angular momentum to the standard Dirac harmonic oscillator where the Dunkl derivative parameter plays the role of the angular momentum eigenvalue. However, unlike the previous case, here the eigenfunction and the energy spectrum explicitly depend on the Dunkl derivative parameter $\mu$. Further, if we try to impose that the spinor components $\Psi_1$ and $\Psi_2$ are simultaneously even or odd functions, mathematically the solutions of the differential equations (\ref{e1}) and (\ref{e2}) are incompatible and erroneous.

The energy spectrum of the two cases studied in this Section can be combined into a single general expression as follows
\begin{equation}
E=\pm mc^2\sqrt{1+\frac{4\hbar\omega}{mc^2}n_{\mu}}, \quad\quad n_{\mu}=n+\left(\frac{\mu}{2}+\frac{1}{2}\right)\left(1-(-1)^n\right),
\end{equation}
where if $n$ is even we get the case in which $\Psi_1$ is an even function and $\Psi$ is an odd function. In the same way, if $n$ is odd, we get the case which $\Psi_1$ is an odd function and $\Psi$ is an even function. This result is in full agreement with that obtained previously in Ref. \cite{Chung}, where the energy spectrum was obtained using the so-called Wigner-Dunkl algebra.

\section{The $SU(1,1)$ Perelomov coherent states and the standard Dirac-Moshinsky oscillator limit}

The purpose of this Section is to construct the $SU(1,1)$ Perelomov coherent states for the $(1+1)$ Dirac-Dunkl oscillator. These coherent states are defined as the action of the displacement operator $D(\xi)$ onto the lowest normalized state $|k,0\rangle$ \cite{Perellibro}
\begin{equation}
|\zeta\rangle=D(\xi)|k,0\rangle=(1-|\zeta|^2)^k\sum_{n=0}^\infty\sqrt{\frac{\Gamma(n+2k)}{n!\Gamma(2k)}}\zeta^n|k,n\rangle.\label{PCN}
\end{equation}
The displacement operator $D(\xi)$ is defined in terms of the $SU(1,1)$ algebra generators $K_+, K_-$ as
\begin{equation}
D(\xi)=\exp(\xi K_{+}-\xi^{*}K_{-}),\label{do}
\end{equation}
where $\xi=-\frac{1}{2}\tau e^{-i\varphi}$, $-\infty<\tau<\infty$ and $0\leq\varphi\leq2\pi$. Thus, if we substitute the Sturmian basis (equation (\ref{gen})) into the equation (\ref{PCN}), we obtain the following result
\begin{equation}
\Psi(r,\zeta)=\left[\frac{2(1-|\zeta|^2)^{2k}}{\Gamma(2k)}\right]^{1/2}r^{2k-\frac{1}{2}}e^{-r^2/2}\sum_{n=0}^\infty\zeta^nL_n^{2k-1}(r^2).
\end{equation}
This sum can be computed if we use the generating function for the Laguerre polynomials
\begin{equation}
\sum_{n=0}^\infty L_n^\nu(x)y^n=\frac{e^{xy/(1-y)}}{(1-y)^{\nu+1}}, \quad\quad |y|<1,\label{lags}
\end{equation}
to obtain the closed form of the $SU(1,1)$ Perelomov coherent states in terms of the group number $k$
\begin{equation}
\Psi(r,\zeta)=\left[\frac{2(1-|\zeta|^2)^{2k}}{\Gamma(2k)}\right]^{1/2}\frac{r^{2k-\frac{1}{2}}}{(1-\zeta)^{2k}}e^{\frac{r^2(1-3\zeta)}{2(\zeta-1)}}.
\end{equation}
Therefore, for the case in which $\Psi_1$ is an even function and $\Psi_2$ is an odd function, the $SU(1,1)$ Perelomov coherent states for the $(1+1)$ Dirac-Dunkl oscillator are given by
\begin{equation}
\Psi_1(r,\zeta)=\left[\frac{2(1-|\zeta|^2)^{\frac{1}{2}+\mu}}{\Gamma\left(\frac{1}{2}+\mu\right)}\right]^{1/2}\frac{r^{\mu}}{(1-\zeta)^{\frac{1}{2}+\mu}}e^{\frac{r^2(1-3\zeta)}{2(\zeta-1)}},
\end{equation}
\begin{equation}
\Psi_2(r,\zeta)=\left[\frac{2(1-|\zeta|^2)^{\frac{3}{2}+\mu}}{\Gamma\left(\frac{3}{2}+\mu\right)}\right]^{1/2}\frac{r^{\mu+1}}{(1-\zeta)^{\frac{3}{2}+\mu}}e^{\frac{r^2(1-3\zeta)}{2(\zeta-1)}}.
\end{equation}
With these results we can write the spinor of the coherent states $\Psi(r,\zeta)$ as
\begin{equation}
\Psi(r,\zeta)=\frac{r^{\mu}}{(1-\zeta)^{\frac{1}{2}+\mu}}e^{\frac{r^2(1-3\zeta)}{2(\zeta-1)}}
\begin{pmatrix}
A\left[\frac{2(1-|\zeta|^2)^{\frac{1}{2}+\mu}}{\Gamma\left(\frac{1}{2}+\mu\right)}\right]^{1/2}\mathbb{I}\\
B\left[\frac{2(1-|\zeta|^2)^{\frac{3}{2}+\mu}}{\Gamma\left(\frac{3}{2}+\mu\right)}\right]^{1/2}\frac{r}{1-\zeta}
\end{pmatrix},
\end{equation}
where $A$ and $B$ are two normalization constants. Moreover, since the final expression of the coherent states only depends on the group number $k$, the spinor of the coherent state $\Psi(r,\zeta)$ for the case when $\Psi_1$ is an odd function and $\Psi_2$ is an even function is
\begin{equation}
\Psi(r,\zeta)=\frac{r^{\mu}}{(1-\zeta)^{\frac{1}{2}+\mu}}e^{\frac{r^2(1-3\zeta)}{2(\zeta-1)}}
\begin{pmatrix}
A'\left[\frac{2(1-|\zeta|^2)^{\frac{3}{2}+\mu}}{\Gamma\left(\frac{3}{2}+\mu\right)}\right]^{1/2}\frac{r}{1-\zeta}\\
B'\left[\frac{2(1-|\zeta|^2)^{\frac{1}{2}+\mu}}{\Gamma\left(\frac{1}{2}+\mu\right)}\right]^{1/2}\mathbb{I}
\end{pmatrix}.
\end{equation}
The closed form of the $SU(1,1)$ Perelomov coherent states of the Dirac-Dunkl oscillator have been obtained due to the generating function for the Laguerre polynomials (equation (\ref{lags})).

Now, we study the case when the parameter of the Dunkl derivative vanishes ($\mu=0$), to recover the standard Dirac-Moshinsky oscillator. To this end, we introduce the generalized Hermite polynomials \cite{GEN1,RoslerH}
\begin{equation}
H_{2m+p}^{\mu}(r)=(-1)^m\sqrt{\frac{\Gamma(m+1)}{\Gamma(m+p+\mu+\frac{1}{2})}}r^pL_m^{\mu-\frac{1}{2}+p}(r^2),
\end{equation}
where $p=0,1$. Thus, we can write the Laguerre polynomials of our problem in terms of the generalized Hermite polynomials as follows
\begin{equation}
H_{2n}^{\mu}(r)=(-1)^n\sqrt{\frac{\Gamma(n+1)}{\Gamma(n+\mu+\frac{1}{2})}}L_n^{\mu-\frac{1}{2}}(r^2), \quad\quad p=0, \quad m=n,
\end{equation}
\begin{equation}
H_{2n-1}^{\mu}(r)=(-1)^{n-1}\sqrt{\frac{\Gamma(n+1)}{\Gamma(n+\mu+\frac{1}{2})}}\frac{r}{\sqrt{n}}L_{n-1}^{\mu+\frac{1}{2}}(r^2), \quad\quad p=1, \quad m=n-1.
\end{equation}
Therefore, for the case in which $\Psi_1$ is an even function and $\Psi_2$ is an odd function we can write the eigenfunction $\Psi(r)$ as follows
\begin{equation}
|\Psi\rangle=\begin{pmatrix}
\sqrt{\frac{E\pm mc^2}{2E}}A_nr^{\mu}e^{-r^2/2}H_{2n}^{\mu}(r^2)\\
\mp i\sqrt{\frac{E\mp mc^2}{2E}}B_nr^{\mu}e^{-r^2/2}H_{2n-1}^{\mu}(r^2)
\end{pmatrix}.
\end{equation}
If we set $\mu=0$ and make the change $2n\rightarrow n$ in this expression, we obtain up to a constant factor that
\begin{equation}
|\Psi\rangle=\begin{pmatrix}
\sqrt{\frac{E\pm mc^2}{2E}}A_ne^{-r^2/2}H_{n}(r^2)\\
\mp i\sqrt{\frac{E\mp mc^2}{2E}}B_ne^{-r^2/2}H_{n-1}(r^2)
\end{pmatrix},\label{finr}
\end{equation}
with the energy spectrum given by
\begin{equation}
E=\pm mc^2\sqrt{1+\frac{2\hbar\omega}{mc^2}n}. \quad\quad n=0,1,2,...
\end{equation}
These are the eigenfunctions and energy spectrum of the standard $(1+1)$ Dirac-Moshinsky oscillator of equations (\ref{spectrum1}) and (\ref{eigen1}).
A similar relationship can be obtained for the case in which $\Psi_1$ is an odd function and $\Psi_2$ is an even function, in which the equation (\ref{fin2}) is reduced to the equation (\ref{finr}), and the energy spectrum reduces to
\begin{equation}
E=\pm mc^2\sqrt{1+\frac{2\hbar\omega}{mc^2}(n+1)}. \quad\quad n=0,1,2,...
\end{equation}
In fact, the Bargmann index of both algebra realizations of equations (\ref{rk01})-(\ref{rk-1}) and (\ref{rk02})-(\ref{rk-2}) reduces to
\begin{equation}
k_1=\frac{1}{4}, \quad\quad\quad k_2=\frac{3}{4}.
\end{equation}
These values are obtained from the $su(1,1)$ Lie algebra realization
\begin{equation}
K_+=\frac{1}{2}a^{\dag^2}, \quad\quad K_-=\frac{1}{2}a^2, \quad\quad K_0=\frac{1}{2}a^{\dag}a+\frac{1}{4}, \quad\quad K^2=-\frac{3}{16}=k(k-1),
\end{equation}
written in terms of the creation $a^{\dag}$ and annihilation $a$ operators of the harmonic oscillator. Here, the space $\mathcal{H}_{1/4}$ consists of the even
harmonic oscillator number states, and the space $\mathcal{H}_{3/4}$ consists of the odd harmonic oscillator number states \cite{Vourdas2}
\begin{equation}
\left|n;\frac{1}{4}\right\rangle\rightarrow \left|2n\right\rangle, \quad\quad\quad\quad \left|n;\frac{3}{4}\right\rangle\rightarrow \left|2n+1\right\rangle.
\end{equation}
The $SU(1,1)$ coherent states for this case are given as follows
\begin{equation}
\left|\zeta,\frac{1}{4}\right\rangle=D(\xi)|0\rangle_n, \quad\quad\quad \left|\zeta,\frac{3}{4}\right\rangle=D(\xi)|1\rangle_n,
\end{equation}
where the state $|\zeta,\frac{1}{4}\rangle$ is known as squeezed vacuum and is a superposition of even number states (it belongs
to the space $\mathcal{H}_{1/4}$). The state $|\zeta,\frac{3}{4}\rangle$ is a superposition of odd number states (it belongs
to the space $\mathcal{H}_{3/4}$).

\section{Concluding remarks}

In the present paper we introduced an $SU(1,1)$ algebraic approach to exactly solve the problem of the one-dimensional Dirac-Dunkl-Moshinsky oscillator. To do this, we changed the derivative in the standard Dirac-Moshinsky oscillator by the Dunkl derivative. As an application of the one-dimensional Dirac-Dunkl-Moshinsky oscillator, we mapped this problem onto the Anti-Jaynes-Cummings model of quantum optics, which will be studied in greater depth in a future work. In order to obtain the parity of the spinor components for the standard Dirac-Moshinsky oscillator and the Dirac-Dunkl oscillator, we have study the Dirac equation of these problems under space reflections. We have shown that for both problems, if we impose that the Dirac equation is invariant under the parity operator, then one of the spinor components must be an even function and the other an odd function. We decoupled the differential equations for each spinor component and studied the cases when one of these functions is even and the other function is odd. In order to solve the differential equations of each spinor component, we introduced two $su(1,1)$ Lie algebra realizations, where one of them is suitable for even functions and the other for odd functions. The eigenfunctions and energy spectrum were obtained by using $su(1,1)$ irreducible representation theory and its Sturmian basis.

In the last Section of our work we constructed the closed form of the $SU(1,1)$ Perelomov coherent states for the Dirac-Dunkl oscillator. Also, we showed that the energy spectrum and the eigenfunctions of the standard Dirac-Moshinsky oscillator are fully recovered if we set the Dunkl parameter to vanish ($\mu=0$). Finally, we point out that in Ref. \cite{Chung} the $(1+1)$ Dirac-Moshinsky oscillator was studied in terms of the Wigner-Dunkl algebra. However, in that work, the authors only obtained the energy spectrum of the problem.

\section*{Acknowledgments}
This work was partially supported by SNI-M\'exico, COFAA-IPN, EDI-IPN, EDD-IPN, SIP-IPN project number $20200225$.


\begin{thebibliography} {99}
\bibitem{Ito}       D. Ito, K. Mori, E. Carrieri, Nuovo Cimento A 51, 1119 (1967).
\bibitem{Cook}      P. A. Cook, Lett. Nuovo Cimento 1, 419 (1971).
\bibitem{Mos}       M. Moshinsky, A. Szczepaniak, J. Phys. A 22, L817 (1989).
\bibitem{Mos2}      C. Quesne, M. Moshinsky, J. Phys. A 23, 2263 (1990).
\bibitem{Delgado}   A. Bermudez, M. A. Martin-Delgado, A. Luis, Phys. Rev. A 77, 063815 (2008).
\bibitem{Sadurni}   E. Sadurn{\'i}, J. M. Torres, T. H. Seligman, J. Phys. A: Math. Theor. 43, 285204 (2010).
\bibitem{Ferkous}   N. Ferkous, A. Bounames, Phys. Lett. A 325, 21 (2004).
\bibitem{Villalba}  V. M. Villalba, Phys. Rev. A 49, 586 (1994).
\bibitem{Benitez}   J. Ben{\'i}tez, R. P. Mart{\'i}nez y Romero, H. N. N\'u\~nez-Y\'epez, A. L. Salas-Brito, Phys. Rev. Lett. 64, 1643 (1990).
\bibitem{Moreno}    M. Moreno, A. Zentella, J. Phys. A: Math. Gen. 22, L821 (1989).
\bibitem{deLima}    R. de Lima Rodrigues, Phys. Lett. A 372, 2587 (2008).
\bibitem{Jaynes}    E. T. Jaynes, F. W. Cummings, Proc. IEEE 51, 89 (1963).
\bibitem{Bermudez}  A. Bermudez, M. A. Martin-Delgado, E. Solano, Phys. Rev. A 76, 041801 (2007).
\bibitem{Mandal}    B. P. Mandal, S. Verma, Phys. Lett. A 374, 1021 (2010).
\bibitem{NOS0}      D. Ojeda-Guill\'en, R. D. Mota, V. D. Granados, Commun. Theor. Phys. 63, 271 (2015).
\bibitem{Sch}       E. Schr\"odinger, Naturwiss. 14, 664 (1926).
\bibitem{Glau}      R. J. Glauber, Phys. Rev. 130, 2529 (1963).
\bibitem{Israel}    Y. Israel, et. al. Optica, 6, 753 (2019).
\bibitem{Bozzio}    M. Bozzio, E. Diamanti, F. Grosshans, Phys. Rev. A 99, 022336 (2019).
\bibitem{Marwah}    A. S. Marwah, N. L\"utkenhaus, Phys. Rev. A 99, 012346 (2019).
\bibitem{Chung2}    W. S. Chung, H. Hassanabadi, Few-Body Syst. 60, 43 (2019).
\bibitem{Chung3}    W. S. Chung, H. Hassanabadi, Eur. Phys. J. Plus 134, 394 (2019).
\bibitem{Chung4}    W. S. Chung, H. Hassanabadi, Int. J. Theor. Phys. 59, 1069 (2020).
\bibitem{YANG}      L. M. Yang, Phys. Rev. 84, 788 (1951).
\bibitem{DUNKL}     C. F. Dunkl, Trans. Am. Math. Soc. 311, 167 (1989).
\bibitem{XuY}       C. F. Dunkl, Y. Xu, \textit{Orthogonal polynomials of several variables, Encyclopedia of Mathematics and Its
                    Applications}, Vol. 81, Cambridge University Press, Cambridge, 2001.
\bibitem{Rosler}    M. R\"osler, \textit{Dunkl operators: theory and applications}. In: Orthogonal Polynomials and Special Functions (Lecture Notes in
                    Mathematics Vol. 1817), Springer, Berlin, 2003.
\bibitem{Brink}     L. Brink, T. H. Hansson, S. Konstein, M. A. Vasiliev, Nucl. Phys. B 401, 591 (1993).
\bibitem{Lapointe}  L. Lapointe, L. Vinet, Commun. Math. Phys. 178, 425 (1996).
\bibitem{GEN1}      V. X. Genest, M. E. H. Ismail, L. Vinet, A. Zhedanov, J. Phys. A: Math. Theor. 46, 145201 (2013).
\bibitem{GEN2}      V. X. Genest, M. E. H. Ismail, L. Vinet, A. Zhedanov, Commun. Math. Phys. 329, 999 (2014).
\bibitem{GEN3}      V. X. Genest, L. Vinet, A. Zhedanov, J. Phys. Conf. Ser. 512, 012010 (2014).
\bibitem{GEN4}      V. X. Genest, A. Lapointe, L. Vinet, Phys. Lett. A 379, 923 (2015).
\bibitem{NOS1}      M. Salazar-Ram{\'i}rez, D. Ojeda-Guill\'en, R. D. Mota, V. D. Granados, Eur. Phys. J. Plus 132, 39 (2017).
\bibitem{NOS2}      M. Salazar-Ram{\'i}rez, D. Ojeda-Guill\'en, R. D. Mota, V. D. Granados, Mod. Phys. Lett. A 33, 1850112 (2018).
\bibitem{Chung5}    W. S. Chung, H. Hassanabadi, Mod. Phys. Lett. A 34, 1950190 (2019).
\bibitem{Chung}     S. Sargolzaeipor, H. Hassanabadi, W. S. Chung, Mod. Phys. Lett. A 33, 1850146 (2018).
\bibitem{NOS3}      R. D. Mota, D. Ojeda-Guill\'en, M. Salazar-Ram{\'i}rez, V. D. Granados, Ann. Phys. 411, 167964 (2019).
\bibitem{Toyama}    Y. Nogami, F.M. Toyama, Can. J. Phys. 74, 114 (1996).
\bibitem{Nogami}    F. M. Toyama, Y. Nogami, F. A. B. Coutinho, J. Phys. A: Math. Gen. 30, 2585 (1997).
\bibitem{Rados}     R. Szmytkowski, M. Gruchowski, J. Phys. A: Math. Gen. 34, 4991 (2001).
\bibitem{Vourdas}   A. Vourdas, Phys Rev. A 41, 1653 (1990).
\bibitem{Gerry}     C. C. Gerry, J. Kiefer, Phys. Rev. A 38, 191 (1988).
\bibitem{Gur}       Y. Gur, A. Mann, Phys. At. Nucl. 68, 1700 (2005).
\bibitem{RoslerH}    M. R\"osler, Commun. Math. Phys. 192, 519 (1998).
\bibitem{Perellibro} A. M. Perelomov, Generalized Coherent States and Their Applications, Springer, Berlin, 1986.
\bibitem{Vourdas2}  A. Vourdas, J. Phys. A: Math. Gen. 39, R65 (2006).

\end{thebibliography}
\end{document}